# Software for Microscopy Workshop White Paper

Raghav Chhetri[1], Stephan Preibisch[1,2], Nico Stuurman[3]

[1] HHMI Janelia Research Campus, Ashburn, VA, USA
[2] Berlin Institute for Molecular Systems Biology, MDC, Berlin, Germany
[3] Dept. of Cellular and Molecular Pharmacology, University of San Francisco and HHMI/UCSF, USA

## Workshop Participants:

**Nenad Amodaj**, Luminous Point
**Hazen Babcock**, Harvard University
**David Bennett**, Janelia Research Campus/HHMI
**Andreas Boden**, KTH Royal Institute of Technology, Stockholm
**Ulrike Boehm**, Janelia Research Campus/HHMI
**Peter Brown**, Arizona State University
**Ahmet Can Solak**, Chan Zuckerberg Biohub
**Yannan Chen**, Columbia University
**Raghav Chhetri**, Janelia Research Campus/HHMI
**Kevin Dean**, UT Southwestern Medical Center
**Michael DeSantis**, Janelia Research Campus/HHMI
**Brian English**, Janelia Research Campus/HHMI
**Paul French**, Imperial College London
**Adam Glaser**, University of Washington
**John Heddleston**, Janelia Research Campus/HHMI
**Elizabeth Hillman**, Columbia University
**Georg Jaindl**, Vidrio Technologies
**Justine Larsen**, Chan Zuckerberg Initiative
**Jeffrey Kuhn**, Massachusetts Institute of Technology
**Sunil Kumar**, Imperial College London
**Xuesong Li**, National Institutes of Health/NIBIB
**Brian Long**, Allen Institute
**Rusty Nicovich**, Allen Institute
**Henry Pinkard**, University of California, Berkeley
**Stephan Preibisch**, Janelia Research Campus/HHMI
**Blair Rossetti**, Janelia Research Campus/HHMI
**Loic Royer**, Chan Zuckerberg Biohub
**Doug Shepherd**, Arizona State University
**Nicholas Sofroniew**, Chan Zuckerberg Initiative
**Elliot Steel**, University of Sheffield
**Nico Stuurman**, University of California, San Francisco
**Mark Tsuchida**, University of Wisconsin
**Nikita Vladimirov**, Berlin Institute of Medical System Biology (BIMSB)
**Fabian Voigt**, University of Zurich
**Chen Wang**, Janelia Research Campus/HHMI
**Richard Yan**, Columbia University
**Eric Wait**, Janelia Research Campus/HHMI
**Andrew York**, Calico Life Sciences LLC
**Ting Zhao**, Janelia Research Campus/HHMI

## Workshop Video Coverage:

Plenary talk by Nico Stuurmann:  https://youtu.be/JEzsy-qtcbE
5-min presentations of all attendees (part 1/2):  https://youtu.be/N55imdegYPc
5-min presentations of all attendees (part 2/2):  https://youtu.be/irAKA4wVf_Y

## Community Resources:

Temporary Community Page: https://github.com/nicost/uScopeControl

# Summary

Microscopes have morphed from purely optical instruments into motorized, robotic machines that form images on digital sensors rather than eyeballs. This continuing trend towards automation and digitization enables many new approaches to microscopy that would have been impossible or impractical without computer interfaces.  Accordingly, today's development of new microscopes most often depends on concurrent software development to bring these custom-build systems to life. This dependence on software brings opportunities and challenges. Most importantly, a key challenge while developing new microscopes is to develop the appropriate software. Despite the fact that software is easily copied and distributed, remarkably few opportunities are available to share experiences creating microscope control software.  In turn, this brings challenges in creating maintainable and flexible software code and writing User Interfaces (UIs) that are easily used by researchers, who are primarily life scientists.

To start to address these challenges by identifying common problems and shared solutions, we assembled a small group of researchers that develop or use software to control their custom-build microscopes at the Janelia Research Campus for a two-day workshop in February 2020.  The outcome of the workshop was the definition of clear milestones, as well as the recognition of an involved community, much larger than the one assembled at the workshop. This community encounters similar hurdles and shares a great desire to overcome these by stronger, community-wide collaborations on Open Source Software.  This White Paper summarizes the major issues identified, proposes approaches to address these as a community, and outlines the next steps that can be taken to develop a framework facilitating shared microscope software development, significantly speeding up development of new microscopy systems.

# The Current Situation

A few different approaches for software control of custom-build microscopes exist:

1. Create software *de novo* by writing the code in industry standard programming languages such as C, C++, C#, Java, or Python. This approach is often taken since it gives greatest flexibility to the developers, and, since the code is owned by the authors, the software can be shared easily with other researchers. This approach involves a considerable amount of effort on the part of the developer. Most participants of the workshop taking this approach use the Python programming/scripting language, which allows for more rapid development than the other languages listed here.

2. Create software using commercial environments that provide toolboxes/libraries for device control and UI building, such as LabView and MATLAB.  A significant number of the participants use LabView-based code as LabView integrates well with the National Instruments digital and analog IO equipment that is widely used to operate scanning light beams and attain hardware level synchronization of microscope equipment.  Maintaining

LabView code appears to be difficult as it does not integrate well with modern source code repositories and the LabView code used by participants was most often written and maintained by a commercial company (e.g. Coleman Technologies now called Sciotex: https://sciotex.com/our-services/software-development/). The use of MATLAB to create UIs for custom applications was mentioned, while those operating custom two-photon laser scanning microscopes rely on ScanImage, an open source software package written in MATLAB.

3. Operate custom-build microscope using the Open Source microscope control software package µManager (https://micro-manager.org) or by writing code that runs in the µManager environment. µManager consists of a hardware abstraction layer written in C++ that can be exported to other environments (currently Java and Python 2.7, work is ongoing for full Python 3 support under the Python package manager *pip*). The UI of µManager was created in close collaboration with users and Cell Biologists generally find the software easy to use.

4. Commercial software packages for microscope control such as Slidebook (3I), Metamorph (Molecular Devices), Zen (Zeiss), NISElements (Nikon). Extensibility of these packages is usually limited, but technical support is available and the UI of these packages is geared towards life sciences researchers. No workshop participants used this approach.

An overwhelming majority of the participants were unhappy with their current approach. General sentiments were that it is often easier to build the hardware of a new microscope system compared to writing the software. Therefore, software development is often seen as a necessary burden to accomplish the goal of creating a new microscope system. In many cases sharing the software code proves difficult due to: 1. licensing issues, 2. code that was written with a very specific microscope system in mind, thus being tightly coupled to specific hardware, 3. a need to use different programming languages, 4. existing code being difficult to understand, maintain and/or extend. For instance, µManager has limitations in its integration of analog/digital Input/Output capabilities, and many present at the meeting opined that the entry barrier to working on the µManager source code is too high. Moreover, certain imaging modalities such as life-time imaging, wavelength scanning, and non-camera-based imaging are more difficult to implement in µManager. Also, scripting in µManager is not as easy as it should be.

A concurrent important issue is the choice of data storage format. In most instances, data needs to be converted to another format before analysis and/or visualization can take place, which is a waste of energy and time. The lack of a common data storage format is partly explained by the different needs of data producers (i.e. the acquisition software) and data consumers (i.e. the image analysis software), but also by lack of consensus and easy to use libraries for imaging data writing and reading. Likewise, there is an increasing number of applications where image analysis feeds back into acquisition (in the simplest case by focussing the sample based on image quality, but much more elaborate feedback of analysis results into the microscope control software is possible and increasingly desirable), however, there is no common interface for analysis feedback, hence the "glue code" coupling analysis and acquisition needs to be rewritten every time a new application is developed.

## What is desired?

At least three distinct groups of experts are involved in custom-build microscopes: optics experts building the microscope hardware, software developers creating the control software and UI, and the scientists using the systems in their research. There is an overarching desire for tools that are easy to use and understand to each of these groups.  This brings the need for modular tools; components with limited scope that can be used in isolation and that are well described and well-documented.  Hardware developers need building blocks that can communicate to all commonly used hardware components, must be able to specify hardware-based synchronization of components, execute different orders of component state changes, acquire images through cameras, point scanners, or small area detectors, view images, and loop analysis results back into the acquisition engine.  Software developers will need to develop, test, document, and support these tools, while the scientists using these tools will provide invaluable feedback to make the control software more versatile and robust. Long term maintainability mandates high standards in coding practices. Scientists using the microscope systems highly benefit from User Interfaces that clearly guide them along the most optimal workflows.  These workflows may be different depending on the microscope system and the specific experiments they are conducting. In addition, they need easy tools to change the workflow, for instance by scripting or visual programming tools.

We are in the midst of a diversification of programming languages used by imaging software tool developers.  Whereas previously much development occurred in Java, currently Python (among others like Julia) is gaining traction.  The choice for Java was partly caused by the prevalence of the freely available, public domain image analysis package ImageJ.  However, many find it easier to develop in Python (even though there are technical obstacles involving for example multithreading in Python).  Therefore, the to-be-developed modular software building blocks for microscope control should be usable from multiple programming languages to ensure longevity of these tools.  Likewise, the ability to build tools on multiple platforms (i.e. Windows, Linux, and Mac OS) - even though driver limitations likely limit operation mostly to the Windows platform - will help future proof developed software code.

As with other parts of the software chain, modular building blocks to store and load data (in memory, on disk, either locally or remotely) are needed.  It was therefore deemed less important what the actual data format(s) look(s) like, but instead the need for well-defined re-usable interfaces for data and metadata writing and reading that will be used to implement well-supported, pluggable libraries for the desired data formats was clearly acknowledged.

The only way for these new tools to be successful is for them to be widely used, and the best way to achieve this is to have strong involvement by the community in setting the goals, steering development and evaluating accomplishments.  Thus, a governance model ensuring strong involvement of developers and users of the software tools is highly desirable.

## Proposed Approaches

### Data saving & loading module

The key concept behind this module is the definition of interfaces that define access to a minimal set of functionalities required for data saving and loading, thereby allowing the use of almost any image data format (e.g. TIFF, HDF5, N5, ZARR, …). These interfaces should be dimension and data type agnostic and include definitions for handling multi-resolution data while being open for future extensions without breaking functionality. This could be achieved by following the tried-and-tested concepts of [ImgLib2](). The design of such a module very much ties into current discussions about interfaces to data storage. Clearly, there is a strong need for highly performant libraries that are tuned to writing data. CZI awarded a grant to the OME team towards the design of a new image data format ([https://forum.image.sc/t/next-generation-file-formats-for-bioimaging/31361](https://forum.image.sc/t/next-generation-file-formats-for-bioimaging/31361)), although it is unclear whether this involves design of general purpose interfaces to image data and metadata.  Coordination with this group will be critical to avoid duplication, and involvement of the community of microscope control software developers will be critical to successful developments. Ideally, efforts will result in a specification of interfaces for reading and writing data as well as a performant reference implementation that are usable from multiple languages (i.e. Python, Java, C/C++) and multi-platform.

### Hardware abstraction module

Many synergies and opportunities for collaboration exist at the level of hardware abstraction.  µManager contains a hardware abstraction layer first designed in 2005 as well as interface code for a large number of devices.  Several Python projects are working towards hardware abstraction (for instance, see: [https://github.com/python-data-acquisition/meta](https://github.com/python-data-acquisition/meta) and [https://github.com/MicronOxford/microscope](https://github.com/MicronOxford/microscope)), but do not support as many hardware components as µManager . There was a clear desire at the workshop to learn from each other and to work towards a unified layer that can be used by everyone.

The µManager hardware abstraction layer (called MMCore) is written in C++ and exported to other languages using [SWIG](). Java wrappers are used in the µManager UI, and bindings for Python 2.7 are included in the µManager binary distribution.  Efforts are almost complete to package [MMCore as a Python 3 library]() that can be installed using the Python package management tool *pip*.  Hardware abstraction in µManager is a mixture of imperative and state-based control.  Each device can declare state-based properties, that can be queried, read and changed through the API. In addition, device classes have their own imperative control that is specific to a device class.  For instance, shutter devices have imperative functions to open and close the shutter, camera devices have functions to start and stop streaming images, etc.. Access to devices always takes place through MMCore, which also provides facilities such as shutter synchronization with camera exposure, a circular buffer for camera images, caching of state properties, grouping of properties, and storage of metadata such as pixel size and affine transform between stage motion and camera pixels. The clean separation into a MMCore bottom API that communicates with devices, and a top API that provides an abstract interface to all microscope equipment, was instrumental in the development of so many device adapters by so many different developers (most of whom had no complete understanding of the µManager software design and quirks).  Python developers at the workshop voiced a strong desire not to have such a layer in between their code and the devices. It will be worthwhile to discuss those design questions further to better understand requirements and what kind of interfaces will be needed to operate without an intermediate layer.

The basic design of MMCore originated in 2005, and certain limitations have become clear since.  For instance, there are different pathways for images that are acquired by "snapping" and that are acquired by streaming, which is undesirable and confusing. MMCore can work with multiple cameras running simultaneously, but only if their image sizes are identical. Image data are copied from the camera driver into the circular buffer, and from there to the "consumer's" memory space, which reduces performance for fast streaming of large images.  These bottlenecks will need to be addressed before MMCore can function as a community platform for device interfacing. For performance reasons, the state of devices is cached in MMCore. Devices that change state autonomously (for instance, because the user presses a button on the device), can signal the state change to MMCore that will update its cache with this new information.  However, this needs complicated device adapter code (the adapter needs to run its own thread in order to provide such updates), and MMCore does not know which devices provide such updates, and which ones do not. Likewise, devices can call back to upstream layers to inform them of state or other changes.  Those callbacks are currently ad-hoc and would benefit from a more unified mechanism.

We have a great opportunity to develop a new device abstraction layer using the lessons learned in the µManager project as well as in the many Python-based projects for microscope control currently being developed.  Hopefully this can be done in such a way that the treasure chest of existing device control code in the µManager source code repository can be reused (either using a compatibility layer or by semi-automated conversion of the code).
One point of concern amongst Python developers was the complexity of developing and debugging µManager device adaptors.  This is due in part to the use of old tool chains (µManager still uses Visual Studio 2010 (C++03)) that can be alleviated by switching to newer development environments, but also to the use of C++, which can be difficult to develop or debug for those who lack experience in this language.  Several participants at the workshop suggested looking into tools that convert Python code in C/C++/machine code, such as numba, or jax. The idea was also floated to make it easy to create device adapters for "simple" devices that follow straight forward command-response communication using XML/YAML/JSON description files that are compiled into adapters.

Simulated hardware, i.e. software devices that behave as much as possible as the real devices will be essential at many stages of creating the infrastructure code, for testing code that uses the device abstraction layer, and for troubleshooting specific problems/bugs (i.e. can the problem be reproduced with simulated hardware).  µManager has simulated hardware devices

(in the "DemoCamera" device adapter), but that code has become overly complex and will benefit from intensive refactoring and rewriting.  In addition, more realistic images and mechanisms to generate images realistic for specific new imaging situations will be immensely useful.

Although it will be ideal if the C++/Java and Python based developers can collaborate on a hardware device abstraction layer, it is already very useful if these two communities can learn from each other and their requirements.  Discussing design matters in a single place (such as started here: https://github.com/python-data-acquisition/meta) will be beneficial to all.

## Synchronization module

On top of a module capable of communicating with hardware devices, a module - named "Acquisition Engine" in µManager - is needed that synchronizes the actions of the devices, based on a protocol specified by the user.  A simple example of such synchronization is opening of the shutter for the illuminating light source just before camera exposure, and closing it immediately after.  Many components can be involved (i.e., Z-stages, XY-stages, filter wheels, galvo mirrors, lasers, etc..), and it is critical that all devices are in the correct state (and not changing/moving) when images are acquired, and that there is as little wait time as possible (which would extend the duration of the experiment and/or duration of illumination of the sample).  Since commonly used computers do not run a real-time Operating System, and because communication to devices can take considerable time, it is often most efficient to synchronize equipment through electronic signals (TTL signals).  The design of electronic synchronization, including the decision which device functions as the "master clock" differs between microscope systems.  Ideally, the Acquisition Engine understands the electronic synchronization layout and can send protocols to all devices and rely on faithful execution.  In addition, timing information should be used for devices that can not be electronically synchronized.

Abstracting all possible microscope configurations into an Acquisition Engine is difficult, and current code most often is hard coded for a specific microscope system.  The µManager Acquisition Engine has limited support for electronic synchronization (the camera is always used as the master clock and "slave" devices are supposed to change state without delay upon receiving an electronic trigger), and has no mechanism to understand device delays.  In addition, the µManager Acquisition Engine is written in the Java Virtual Machine compatible programming language "Clojure", drastically restricting the number of developers who can work on that code.  Building a new Acquisition Engine that understands multiple electronic synchronization schemes, yet is easy to understand and modify is a useful, yet difficult task.

Most current custom microscopes use National Instruments (NI) equipment for electronic synchronization.  Often combined with wave-form generation for galvo mirrors, these devices either function as the master clock, sending pulses to devices at specified delays to initiate specific actions, or execute actions upon receiving a trigger.  NI equipment is most easily programmed from the non Open Source LabView environment, which is one of the reasons many custom build microscopes are operated using LabView.  Supporting multiple versions of older NI equipment - especially ensuring the equipment responds correctly to input triggers - was so problematic that Vidrio (https://vidriotechnologies.com/) - the company around the Open Source microscope acquisition platform ScanImage) is now developing its own digital/analog IO hardware.  Several other more or less open alternatives exist (for instance: https://doi.org/10.1038/s41598-019-48455-z, http://arc.austinblanco.com/product/triggerscope-3/). Finding a manageable path forward in this realm will be of considerable interest.

## Graphical User Interface

The GUI needs to display images so that the user can arrange the experiment correctly (by, for instance, finding the object of interest), and to monitor progress.  Also, the UI should let the user specify the desired experimental protocol. A plugin mechanism is desirable, so that extensions can be easily used.   The GUI should be targeted to researchers using microscopes rather than microscope builders (although specific UIs for builders may be appropriate).

The most complicated part of the GUI is the viewer, which should be capable of displaying continuously updating live streams efficiently, display overlays of multi-channel, time-lapse, multi-position, spectral, and life-time images, control brightness/contrast/gamma, have live histogram displays, line profiles, tools for region of interest (ROI) selection, 3D display, and facilities for (z) projections.  The µManager viewer, written in Java using the Swing framework, has most of these facilities, but the code is not easily separable into modules, and depends on ImageJ code for the actual display and ROI tooling. The viewer Napari (https://github.com/napari/napari), written in Python using the Qt framework, is actively developed and still in the alpha stage, and may soon be usable as a viewer for live data.  Another Java, ImgLib2-based alternative is BigDataViewer (https://imagej.net/BigDataViewer) that is capable of interactively displaying very large datasets.
Any GUI should be developed in close collaboration with scientists using the application on a daily basis and user feedback should be incorporated in the design as much as possible (as was done for the µManager GUI).

## Community building, governance, and collaboration

The workshop participants expressed a strong desire to continue working on microscope control software as a community.  First steps should be to reach out to a much wider group then could be physically accommodated during the meeting.  As the skewed gender balance amongst the Workshop participants made clear, much effort should be put towards including under-represented groups.  Community discussions can take place on the discourse based forum https://image.sc, (for higher level and community oriented subjects) or through issues in a github repository (for more technical, directly code-related subjects).  As one of the first steps, a governance structure will need to be put in place, where existing projects such as ImageJ and Napari can serve as examples.  This community will be geared towards microscope control software written in Python/C++/Java, but be open and welcoming to other languages and approaches.  It should be a place where everyone, especially junior developers and microscope builders, feel welcome and valued, and will be able to make contributions. Industry representatives should be able to participate at all levels, and outreach to industry partners (starting with existing connections) should be a high priority.

Industry collaboration is especially important with respect to hardware control. Most microscope hardware is procured from commercial companies, hence the microscope control software interfaces with a driver or communication protocol designed and supported by a commercial company. The best possible outcome would be for the common hardware abstraction layer (outlined above), to become an industry-wide standard supported by companies providing code interfacing their equipment directly to the abstraction layer. Some of this is already happening in the µManager project: a majority of the device adapter code is written and maintained by the companies making the hardware, and several companies write software code that makes use of multiple aspects of µManager. Clearly, if the community convenes around a common hardware interface, the participation level by industry will increase. Early input by our industry partners will be extremely valuable, and we should promote industry participation from the outset.

## What are the next steps?

- **Set up online meetings and collaboration space**

- **A new organisation for unified future of device layer**
    - Identify stakeholders, end-users, hardware developers, software developers, vendors
    - Specify an organizational, contribution and governance model.
    - Identify name / brand for device layer (stick with µManager for now?)
    - Determine fiscal sponsor / foundation that can be connected to (Global Bioimaging, NumFocus)
    - Establish language support
    - Clarify relations with vendors, companies
    - Consider quality criterion / guidelines for device adaptors

- **µManager refactor**
    - µManager joins image.sc forum as a community partner
    - Pull out MMCore + MMDevices (C++ code) into their own GitHub repo
    - Make MMCore easily installable (including pip) / easily testable with CI independent of MM java code / GUI (ongoing in pymmcore repo on github)
    - Provide access to the complete Micro-Manager API in Python using a remote procedure call mechanism (implemented using ZeroMQ by Henry Pinkard and added to Micro-Manager source repository early March 2020, use Pygellan package to use from Python).
    - Work on better simulators within Micro-Manager
    - Improve accessibility documentation focused on device adaptor APIs
    - Expose list of conceptual devices that MM has, look at how it matches current hardware.
    - Investigate adding new devices around waveform generation / AO for light sheet / laser scanning use cases

- o   Investigate Python -> C++ conversion for allowing device adaptors written in python to be used in MMCore

- **Python device adaptor API standardization**
    - o   Unify existing python adaptors for same devices, AQC4, other efforts under one GitHub repository.
    - o   Design API that python community can share including both pure python and C++ code
    - o   Investigate compatibility of this API with existing MM devices

- **Implementation of the data saving & loading modules**

## Phase 2 for a unified future:

- Bring together MM C++ device layer, Java and Python APIs, learning from LabView approach, into new API for an open source device layer for the future
    - o   What does backwards compatibility layer look like for current MM devices
    - o   More advanced devices
    - o   More device coordination, better synchronization